\documentclass[12pt]{iopart}
\begin{document}
\title{ISW effect in Unified Dark Matter Scalar Field 
Cosmologies: an analytical approach}

\author{Daniele Bertacca}
\address{Dipartimento di Fisica ``Galileo Galilei", Universit\'a di Padova, 
and INFN Sezione di Padova, via F. Marzolo, 8 I-35131 Padova Italy}
\ead{daniele.bertacca@pd.infn.it}
\author{Nicola Bartolo}
\address{Dipartimento di Fisica ``Galileo Galilei", Universit\'a di Padova, 
and INFN Sezione di Padova, via F. Marzolo, 8 I-35131 Padova Italy}
\ead{nicola.bartolo@pd.infn.it}

\date{\today}

\begin{abstract}
We perform an analytical study of the Integrated Sachs-Wolfe (ISW) effect within the framework of Unified Dark Matter models based on a scalar field 
which aim at a unified description of dark energy and dark matter. Computing the temperature power spectrum of the Cosmic Microwave Background anisotropies 
we are able to isolate those contributions that can potentially lead to strong deviations from the usual ISW effect occurring in a $\Lambda$CDM universe. 
This helps to highlight the crucial role played by the sound speed in the unified dark matter models. 
Our treatment is completely general in that all the results depend only on the speed of sound of the dark 
component and thus it can be applied to a variety of unified models, including those which are not described by a scalar field but relies on a single 
dark fluid.

\end{abstract}
\maketitle

\section{Introduction}
Observations of large scale structure, search for Ia supernovae (SNIa),
 measurements of the Cosmic Microwave Background (CMB) anisotropies 
all suggest that two dark components govern the dynamics of the universe.
In particular they are the dark matter (DM), responsible for
structure formation, and an additional dark energy component that drives  
the cosmic acceleration observed at present~\cite{perlmutter,riess}. Two main routes have been followed to provide a plausible 
realization of the dark energy, a non-zero cosmological constant $\Lambda$ 
(see, e.g. Ref.~\cite{Weinberg:1988cp}), or a dynamical dark energy (DE) component in the form of a scalar field, like for example 
\emph{Quintessence} \cite{Wetterich88,peebles,earlyqu,Carroll:1998zi,Ferreira97,CLW,caldwell98,Paul99} 
and \emph{k-essence} models \cite{Chiba:1999ka,AMS1,AMS2}. The latter are characterized by a Lagrangian with non-canonical 
kinetic term and inspired by earlier studies of k-inflation \cite{kinf} \cite{Garriga:1999vw} 
(a complete list of dark energy models can be found in the recent review \cite{Copeland:2006wr}). At different levels all of these scenarios 
suffer of non trivial fine tuning problems and of the so called cosmic coincidence problem (why $\Omega_{DM}$   
and $\Omega_{\Lambda}$ are both of order unity today). 

More recently the alternative hypothesis of unified models of dark energy and dark matter has been considered. In this case   
a single fluid behaves both as dark matter and dark energy. This has been variously 
referred to as ``Unified Dark Matter" (UDM), or ``Quartessence". Among several models of k-essence considered in the literature there 
exist several UDM models. First of all \emph{purely kinetic models}, i.e. the 
\emph{generalized Chaplygin gas} (GCG) \cite{Kamenshchik:2001cp,Bilic02,Bento:2002ps}
model, the Scherrer and generalized  Scherrer solutions
\cite{Scherrer:2004au} \cite{Bertacca:2007ux}, a single dark 
perfect fluid with a simple 2-parameter barotropic equation of state
\cite{Quercellini:2007ht}, or the homogeneous scalar field deduced 
from the galactic halo space-times \cite{DiezTejedor:2006qh}.
Moreover there are models in which imposing the Lagrangian of the scalar field
 to be a constant allows directly to describe a unified dark matter/ dark energy fluid
\cite{Bertacca:2007ux} \cite{Gorini} (see also, with different approach,
\cite{Padmanabhan:2002sh}). Alternative approaches to the unification of DM 
and DE have been proposed in Ref.~\cite{takayana}, 
in the frame of supersymmetry, and in Ref.~\cite{bono}, 
in connection with the solution of the 
strong CP problem. 

The recent attitude in analysing the observational consequences of the DE models has been that of considering not only the background equation of 
state and its evolution with time, but also to focus on the sound speed which regulates the growth of the dark energy fluid 
perturbations on different cosmological scales. In this case the sound speed has been often 
treated as a completely independent parameter in order to explore the consequences on the CMB anisotropies and its effects on the low $\ell$ multipoles
~\cite{WL,DD,BO}. The efficiency of this method relies on the observation that, for a single scalar field with canonical kinetic term, the speed of sound is 
equal to the speed of light, and thus it can cluster only on scales of the horizon size, while for other models it can be lower than unity, 
implying the possibility of clustering on smaller scales~\cite{Huss}. Another important issue is whether the dark matter clustering is influenced by 
the dark energy and, for the unified models, it becomes especially relevant in view of this approach. 
In the GCG model (both as dark energy and unified dark matter) strong constraints come from the CMB 
anisotropies~\cite{BDG,Carturan:2002si,Amendola:2003bz}  and the analysis of the mass power spectrum \cite{Sandvik:2002jz}. 
In the Scherrer solution the parameters of the model have to
be fine-tuned in order for the model not to exhibit finite pressure
effects in the non-linear stages of structure formation \cite{Giannakis-Hu}. 

In this paper we consider cosmological models where dark matter and dark
energy are manifestations of a single scalar field, and we focus on the contribution to the 
 large-scale CMB anisotropies which is due to the evolution in time of the gravitational potential
from the epoch of last scattering up now, the so called late Integrated Sachs-Wolfe (ISW) effect~\cite{Sachs:1967er}. Through an analytical approach 
we point out the crucial role of the speed of sound in the unified dark matter models in determining strong deviations from the usual standard ISW 
occurring in the $\Lambda$CDM models. Our treatment is completely general in that all the results depend only on the speed of sound of the dark 
component and thus it can be applied to a variety of models, including those which are not described by a scalar field but relies on a single perfect 
dark fluid. In the case of $\Lambda$CDM models the ISW is dictated by the background evolution, which causes the late time decay of the gravitational 
potential when the cosmological constant starts to dominate~\cite{Hu:1995em}. 
In the case of the unified models there are two simple but important aspects: first, the fluid which 
triggers the accelerated expansion at late times is also the one which has to cluster in order to produce the structures we see today. Second, 
from the last 
scattering to the present epoch, the energy density of the universe is dominated by a single dark fluid, and therefore the gravitational potential 
evolution 
is 
determined by the background and perturbation evolution of just such a fluid. As a result the general trend is that the possible appearance of a sound speed 
significantly different from zero at late times corresponds to the appearance of a Jeans length (or a sound horizon) 
under which the dark fluid does not cluster any more, causing a strong evolution in time of the gravitational potential 
(which starts to oscillate and decay) and thus a strong ISW effect. Our results 
show explicitly that the CMB temperature power spectrum $C_{\ell}$ 
for the ISW effect contains some terms depending on the speed of sound which give a high contribution
along a wide range of multipoles $\ell$. As the most straightforward way to avoid these critical terms one can require the sound speed to be always 
very close to zero (thou see Sec.~3.2.3  for a more detailed discussion on this point). 
Moreover we find that such strong imprints from the ISW effect comes primarily from the evolution 
of the dark component perturbations, rather than from the background expansion history.   
The paper is organized as follows. In Sec.~2 we obtain the evolution equation for the gravitational potential. In Sec.~3 we 
start the analytical analysis 
of the ISW effect, dividing the resulting expression for the angular CMB power spectrum according to three relevant regions: 
those perturbation 
modes that enter the horizon after the acceleration of the universe becomes relevant, 
and perturbation modes that are inside or outside the sound horizon of the dark fluid. 
In Sec.~3.2.2 
we point out those contributions to the ISW effect that are triggered by the sound speed and that are responsible for a strong ISW imprint. Sec.~4
contains our conclusions and a discussion of our results applied to various unified dark matter models. 
\section{Linear perturbations in scalar field unified dark matter models }
We consider the action that describes most of the dark matter unified models within the framework of k-essence 
\begin{equation}\label{eq:action}
S = S_{G} + S_{\varphi} = 
\int d^4 x \sqrt{-g} \left[\frac{R}{2}+\mathcal{L}(\varphi,X)\right] 
\end{equation}
where 
\begin{equation}\label{x}
X=-\frac{1}{2}\nabla_\mu \varphi \nabla^\mu \varphi \;.
\end{equation}
We use units such that $8\pi G = 1$ and signature $(-,+,+,+)$.

The energy-momentum tensor of the scalar field $\varphi$ is
\begin{equation}
  \label{energy-momentum-tensor}
  T^{\varphi}_{\mu \nu } = 
- \frac{2}{\sqrt{-g}}\frac{\delta S_{\varphi }}{\delta
    g^{\mu \nu }}=\frac{\partial \mathcal{L}
(\varphi ,X)}{\partial X}\nabla _{\mu }\varphi
  \nabla _{\nu }\varphi +\mathcal{L}(\varphi ,X)g_{\mu \nu }.
\end{equation}
If $X$ is time-like $S_{\varphi}$ describes a perfect fluid with 
$T^{\varphi}_{\mu \nu }=(\rho +p) u_{\mu} 
u_{\nu }+p\,g_{\mu \nu }$, with pressure 
\begin{equation}
  \label{pressure}
  \mathcal{L}=p(\varphi ,X)\;,  
\end{equation}
and energy density
\begin{equation}
  \label{energy-density}
  \rho =\rho 
(\varphi ,X)= 2X\frac{\partial p(\varphi ,X)}
  {\partial X}-p(\varphi ,X)  
\end{equation}
where 
\begin{equation}
  \label{eq:four-velocity}
  u_{\mu }= \frac{\nabla _{\mu }\varphi }{\sqrt{2X}}.
\end{equation}
 the four-velocity.

Now we assume a flat, homogeneous Friedmann-Robertson-Walker 
background metric i.e.
\begin{equation}
ds^2=-dt^2+a(t)^2\delta_{ij} dx^i dx^j = 
a(\eta)^2 (-d\eta^2+\delta_{ij} dx^i dx^j)\, ,
\end{equation}
where $a(t)$ is the scale factor,$\delta_{ij}$ denotes the unit tensor and $d\eta=dt /a$ is the conformal time. 
In such a case, the background evolution of the universe is characterized
completely by the following equations 
\begin{equation}
\label{eq_u1}
\mathcal{H}^2=a^2 H^2 =\frac{1}{3} a^2 \rho\, ,
\end{equation}
\begin{equation}
\label{eq_u2}
\mathcal{H}'-\mathcal{H}^2=a^2 \dot{H} = - \frac{1}{2} a^2(p + \rho)\, ,
\end{equation}
where ${\cal H}=a'/a$, the dot denotes differentiation w.r.t. the cosmic time $t$ and 
a prime w.r.t. the conformal time $\eta$.
On the background $X=\frac{1}{2}\dot{\varphi}^2=\varphi'^2/(2a^2)$ and 
the equation of motion for the homogeneous mode $\varphi(t)$ becomes
\begin{equation}
 \label{eq_phi}
\left(\frac{\partial p}{\partial X} 
+2X\frac{\partial^2 p}{\partial X^2}\right)\ddot\varphi
+\frac{\partial p}{\partial X}(3H\dot\varphi)+
\frac{\partial^2 p}{\partial \varphi \partial X}\dot\varphi^2
-\frac{\partial p}{\partial \varphi}=0 \;. 
\end{equation}
One of the relevant quantities for the dark energy issue  is the equation of state $w \equiv p/\rho$ which in our case reads 
\begin{equation}
\label{w}
w = \frac{p}{2X \frac{\partial p}{\partial X} - 
p} \;. 
\end{equation} 
On the other hand we will focus on the other relevant physical quantity, the speed of sound,  
which enters in governing the evolution of the scalar field perturbations. Considering small inhomogeneities of the scalar field 
\begin{equation}
\varphi(t,x)= \varphi_0(t)+\delta\varphi(t,\mbox{\boldmath $x$})\, ,
\end{equation}
we can write the metric in the longitudinal gauge as 
\begin{equation}
ds^2=-(1+2 \Phi)dt^2+(1-2 \Phi)a(t)^2\delta_{ij} dx^i dx^j 
\end{equation}
since $\delta T_{i}^j = 0$ for $i \neq j$~\cite{Mukhanov:1990me}.

From the linearized $(0-0)$ and $(0-i)$ Einstein equation one obtains (see Ref.~\cite{Garriga:1999vw}
and Ref.~\cite{Mukhanov:2005sc})
\begin{equation}
\label{pertu-eq1}
\nabla^2 \Phi = \frac{1}{2} \frac{a^2(p+\rho)}{c_s^2 \mathcal{H}} \left(
\mathcal{H}  \frac{\delta\varphi}{\varphi_0'}+ \Phi \right)'\, ,
\end{equation}
\begin{equation}
\label{pertu-eq2}
\left(a^2 \frac{\Phi}{\mathcal{H}} \right)'= \frac{1}{2} \frac{a^2(p+\rho)}
{\mathcal{H}^2}\left(\mathcal{H}  \frac{\delta\varphi}{\varphi_0'}
+ \Phi \right) \; ,
\end{equation}
where one defines a ``speed of sound'' $c_s^2$ relative to the pressure and energy density fluctuation of 
the kinetic term~\cite{Garriga:1999vw} as 
\begin{equation}
\label{cs}
c_s^2 \equiv  \frac{(\partial p /\partial X)}
{(\partial \rho /\partial X)} = 
\frac{\frac{\partial p}{\partial X}}{\frac{\partial p}
{\partial X}+ 2X\frac{\partial^2 p}{\partial X^2}} \;. 
\end{equation}
Eqs.~(\ref{pertu-eq1}) and~(\ref{pertu-eq2}) are sufficient to determine the gravitational potential
$\Phi$ and the perturbation of the scalar field.
Defining two new variables
\begin{equation}
\label{u-v}
u\equiv 2 \frac{\Phi}{(p+\rho)^{1/2}} \;, \quad\quad v\equiv z \left(
\mathcal{H}  \frac{\delta\varphi}{\varphi_0'}+ \Phi \right)\;,
\end{equation}
where $z=a^2(p+\rho)^{1/2}/(c_s \mathcal{H})$,
we can recast (\ref{pertu-eq1}) and (\ref{pertu-eq2}) in terms 
of $u$ and $v$ \cite{Mukhanov:2005sc}
\begin{equation}
\label{pertu-eq_uv}
c_s \triangle u = z \left( \frac{v}{z}\right)'\;,\quad \quad 
c_s v= \theta \left( \frac{u}{\theta}\right)'
\end{equation}
where $\theta = 1/(c_s z)=(1+p/\rho)^{-1/2}/(\sqrt{3}a)$.
Starting from (\ref{pertu-eq_uv}) we arrive at the following second 
order differential equations for $u$ \cite{Mukhanov:2005sc}
\begin{equation}
\label{diff-eq_u}
u''-c_s^2 \nabla^2 u - \frac{\theta''}{\theta}u=0\;.
\end{equation}
Notice that this equation can also be used to describe any perfect fluid 
with equation of state $p=p(\rho)$, up to  a redefinition of $c_s$.
In this case $c_s^2=p'/\rho'$ corresponds to the usual adiabatic sound speed. 
In this way, with the same equation (\ref{diff-eq_u}), we can also describe the $\mathrm{\Lambda CDM}$ model. Also pure kinetic 
Lagrangian (\ref{pressure})  $\mathcal{L}(X)$ models 
(see for example Ref.~\cite{Bertacca:2007ux}), can be described as a perfect fluid 
with the pressure $p$ uniquely determined by the energy density, since they both depend on a single degree of freedom, the 
kinetc term $X$. 

Unfortunately we do not know the exact solution for a generic Lagrangian. However we can consider the asymptotic solutions i.e. 
long-wavelength and short-wavelength perturbations, depending whether $c_s^2k^2 \ll \left|\theta''/\theta\right|$ or 
$c_s^2k^2 \gg \left|\theta''/\theta\right|$, respectively. This means to consider perturbations on scale much larger or 
much smaller than the effective Jeans length for the gravitational potential $\lambda^2_J = c_s^2 \left|\theta/\theta''\right|$.
  
For a plane wave perturbation $u \propto u_k(\eta) 
\exp(i \mbox{\boldmath $k x$})$ in the short-wavelength limit 
($c_s^2k^2 \gg \left|\theta''/\theta\right|$) we obtain
\begin{equation}
\label{u-cs_k>>theta''/theta}
u_k\simeq\frac{C_k(\bar{\eta})}{c_s^{1/2}(\eta)}
\cos \left( k \int_{\bar{\eta}}^{\eta} c_s d\tilde{\eta}\right)\;,
\end{equation}
where $C_k$ is a constant of integration. Instead, neglecting
the decaying mode, the long-wavelength solution
($c_s^2k^2 \ll \left|\theta''/\theta\right|$) is
\begin{equation}
\label{u-cs_k<<theta''/theta}
u_k=A_k(\bar{\eta}) \theta \int_{\bar{\eta}}^{\eta}
\frac{d\tilde{\eta}}{\theta^2}\, ,
\end{equation}
where $A_k$ is a constant of integration .\\
Once $u$ is computed we can obtain the value of the gravitational potential $\Phi$ through Eq.~(\ref{u-v})
and the perturbation of the scalar field from Eq.~(\ref{pertu-eq2})
\begin{equation}
\label{deltaphi}
\delta\varphi=2\sqrt{2X}\frac{\left(\Phi' +\mathcal{H} \Phi \right)}
{a (p+\rho)} \; .
\end{equation}

\section{Analytical approach to the ISW effect}
Let us now focus on the ISW effect. The ISW contribution to the CMB power spectrum is given by
\begin{equation}
\label{Cl}
\frac{2l+1}{4\pi}C_{l}^{\mathrm{ISW}}= \frac{1}{2\pi^2}\int_0^\infty 
\frac{dk}{k}
k^3  \frac{\left|\Theta_{l}^{\mathrm{ISW}}(\eta_0,k)\right|^2}{2l+1}\, ,
\end{equation}
where $\Theta_{l}^{\mathrm{ISW}}$ is the fractional temperature perturbation 
due to ISW effect 
\begin{equation}
\label{theta}
\frac{\Theta_{l}^{\mathrm{ISW}}(\eta_0,k)}{2l+1}=2\int_{\eta_*}^{\eta_0}
\Phi'(\tilde{\eta},k) j_l[k(\eta_0-\tilde{\eta})]d\tilde{\eta}\, ,
\end{equation}
with $\eta_0$ and $\eta_*$ the present and the last scattering 
conformal times respectively and $j_l$ are the spherical Bessel functions.
\\
We now evaluate analytically the power spectrum~(\ref{Cl}). As a first step, following the 
same procedure of Ref.~\cite{Hu:1995em}, we notice that, 
when the acceleration of the universe begins to be important, the 
expansion time scale $\eta_{1/2}=\eta(w=-1/2)$ sets a critical wavelength 
corresponding to $k \eta_{1/2}=1$. It is easy to see that if we consider the $\mathrm{\Lambda CDM}$ model then 
$\eta_{1/2}=\eta_\Lambda$ i.e. when $a_\Lambda/a_0=(\Omega_0/\Omega_\Lambda)
^{1/3} $ \cite{Hu:1995em}. Thus at this critical point we can break the integral (\ref{Cl}) in two parts~~\cite{Hu:1995em}
\begin{equation}
\label{Cl2}
\frac{2l+1}{4\pi}C_{l}^{\mathrm{ISW}}= \frac{1}{2\pi^2}\left[I_{\Theta_{l}}
(k \eta_{1/2} <1) + I_{\Theta_{l}}(k \eta_{1/2} > 1)\right]\, ,
\end{equation}
where
\begin{equation}
\label{I_theta_k-eta<1}
I_{\Theta_{l}}
(k \eta_{1/2} <1) \equiv \int_0^{1/\eta_{1/2}} 
\frac{dk}{k}k^3  \frac{\left|\Theta_{l}^{\mathrm{ISW}}(\eta_0,k)\right|^2}{2l+1}\, ,
\end{equation}
and
\begin{equation}
\label{I_theta_k-eta>1}
I_{\Theta_{l}}(k \eta_{1/2} > 1) \equiv \int_{1/\eta_{1/2}}^\infty 
\frac{dk}{k}
k^3  \frac{\left|\Theta_{l}^{\mathrm{ISW}}(\eta_0,k)\right|^2}{2l+1}\;.
\end{equation}
As explained in Ref.~\cite{Hu:1995em} the ISW integrals (\ref{theta}) 
takes on different forms in these two regimes 
\begin{eqnarray}
\label{theta-approx}
\frac{\Theta_{l \; \mathrm{ISW}}(\eta_0,k)}{2l+1} = 
\left\{ \begin{array}{ll}
2\Delta\Phi_k \; j_l[k(\eta_0-\eta_{1/2})] & k \eta_{1/2} \ll 1\\
2\Phi_k'(\eta_k) I_l / k & k \eta_{1/2} \gg 1
\end{array} \right.
\end{eqnarray}
where $\Delta  \Phi_k$ is the change in the potential from the matter-dominated
(for example at recombination) to the present epoch $\eta_0$ and 
$\eta_k \simeq \eta_0 - (l+1/2)/k $ is the conformal time when a given k-mode 
contributes maximally to the angle that this scale subtends on the sky, obtained at the peak of the Bessel fucntion $j_\ell$. 
The first limit in Eq.~(\ref{theta-approx}) is obtained by approximating the Bessel function as a constant evaluated at the 
critical epoch $\eta_{1/2}$. Since it comes from perturbations of wavelenghts 
longer than the distance a photon can travel during the the time $\eta_{1/2}$, a kick ($2\Delta\Phi_k$) to the photons is the main 
result, and it will corresponds to very low multipoles, since $\eta_{1/2}$ is very close to the present epoch $\eta_0$. It thus 
appears similar to a Sachs-wolfe effect (or also to the early ISW contribution). The second limit in 
 Eq.~(\ref{theta-approx}) is achieved by considering the strong oscillations of the Bessel functions in this regime, and thus 
evaluating the time derivative of the potentials out of the integral at the peak of the Bessel function, leaving the 
integral ~\cite{Hu:1995em} 
\begin{equation}
I_l \equiv \int_0^\infty  j_l(y) dy = \frac{\sqrt{\pi}}{2} 
\frac{\Gamma[(l+1)/2]}{\Gamma[(l+2)/2]}\;.
\end{equation}
With this procedure, replacing (\ref{theta-approx}a) in (\ref{I_theta_k-eta<1})
 and (\ref{theta-approx}b) in (\ref{I_theta_k-eta>1}) we can obtain the 
ISW contribution to the CMB anisotropies power spectrum (\ref{Cl}).

Now we have to calculate, through Eqs.~(\ref{u-cs_k>>theta''/theta})-(\ref{u-cs_k<<theta''/theta}) and 
(\ref{u-v}), the value of $\Phi(k,\eta)$ for $k \eta_{1/2} \ll 1$ and $k \eta_{1/2} \gg 1$.
As we will see tha main differences (and the main difficulties) of the unified dark matter models  
with respect to the $\Lambda$CDM case will appear from the second regime of Eq.~(\ref{theta-approx}).

\subsection{Derivation of $I_{\Theta_{l}}$ for modes $k \eta_{1/2} <1$}

In the UDM models when $k \eta_{1/2} \ll 1$ then $c_s^2k^2 \ll \left|\theta''/\theta\right|$ 
is always satisfied. This is due to the fact that before the dark fluid start to be relevant as a cosmological constant, 
for $\eta < \eta_{1/2}$, its sound speed generically is very close to zero in order to guarantee enough structure formation, and 
moreover the limit  $k \eta_{1/2} \ll 1$ involves very large scales (since $\eta_{1/2}$ is very close to the present epoch). 
For the standard $\Lambda$CDM model the condition is clearly satisifed.  
In this situation we can use the relation (\ref{u-cs_k<<theta''/theta}) 
and $\Phi_k$ becomes
\begin{equation}
\label{Phi_k-eta<1}
\Phi_k=A_k \left(1-\frac{\mathcal{H}(\eta)}{a^2(\eta)} \int_{\eta_i}^{\eta}
a^2(\tilde{\eta})d\tilde{\eta}\right)\, .
\end{equation}
We immediately see that $A_k = \Phi_k(0)$, the large scale gravitational 
potential during the radiation dominated epoch.
The integral in Eq.~(\ref{Phi_k-eta<1}) may be written as follows
\begin{equation}
\label{Ir}
\int_{\eta_i}^{\eta}a^2(\tilde{\eta})d\tilde{\eta}=
I_R + \int_{\eta_R}^{\eta} a^2(\tilde{\eta})d\tilde{\eta}\, ,
\end{equation}
where $I_R=\int_{\eta_i}^{\eta_R}a^2(\tilde{\eta})d\tilde{\eta}$ and 
$\eta_R$ is the conformal time at recombination. When $\eta_i < \eta < \eta_R$
the UDM Models behave as dark matter~\footnote{In fact the Scherrer~\cite{Scherrer:2004au} and generalized 
Scherrer solutions \cite{Bertacca:2007ux} in the very early universe, much before the equality epoch, 
have $c_s \neq 0$ and $w>0$. However  at these times the dark fluid contribution is  
sub-dominant with respect to the radiation energy density and thus ther is no substantial effect on the following equations.}. 
In this temporal range the universe is dominated by a mixture of ``matter'' and radiation and $I_R$
becomes
\begin{equation}
I_R = \eta_* a_{eq} \left(\frac{\xi_R^5}{5}+\xi_R^4+\frac{4\xi_R^3}{3} \right)
\end{equation}
where $a_{eq}$ is the value of the scalar factor at matter-radiation equality,
$\xi=\eta/\eta_*$ and $\eta_*=(\rho_{eq} a_{eq}^2/24)^{-1/2}=\eta_{eq}/
(\sqrt{2}-1)$.  With these definitions it is easy to see that 
$a_R=a_{eq}(\xi_R^2+2\xi_R)$. Notice that Eq.~(\ref{Phi_k-eta<1}) is obtained in the case of adiabatic perturbations. 
Since we are dealing with unified dark matter models based on a scalar field, there will always be an intrinsic non-adiabatic 
pressure (or entropic) pertubation. However for the very long wavelenghts, $k\eta_{1/2}  \ll 1$ under consideration here such an 
intrinsic perturbation turns out to be negligible~\cite{Garriga:1999vw}. For adiabatic perturbations $\Phi_k(\eta_R) \cong (9/10)\Phi_k(0)$ 
\cite{Mukhanov:1990me} and accounting for the primordial power spectrum, $k^3 |\Phi_k(0)|^2=B k^{n-1}$, where $n$ is the scalar 
spectral index, we get from Eq.~(\ref{theta-approx}a)
\begin{eqnarray}
\label{I_theta_k-eta<1-2}
I_{\Theta_{l}}
(k \eta_{1/2} <1) & \approx & 4(2l+1) B \int_0^{1/\eta_{1/2}} 
\frac{dk}{k} k^{n-1} j_l^2[k(\eta_0-\eta_{1/2})] \nonumber \\
& \times & \left| \frac{1}{10} -\frac{\mathcal{H}(\eta_0)}{a^2(\eta_0)}\left[ 
\int_{\eta_R}^{\eta_0} a^2(\tilde{\eta})d\tilde{\eta} \right] \right|^2 \; , 
\end{eqnarray}
where we have neglected $I_R$ since it gives a negligible contribution. 

A first comment is in order here. There is a vast class of  unified dark matter models that are able to reproduce 
exactly the same background expansion history of the universe as the $\Lambda$CDM model (at least from the recombination epoch onwards).
For example this is the case of the the Scherrer and generalized  Scherrer unified models \cite{Scherrer:2004au} \cite{Bertacca:2007ux}, 
the generalized Chaplygin gas~\cite{Kamenshchik:2001cp,Bilic02,Bento:2002ps} for the parameter $\alpha$ which tends to zero, 
the models proposed in Ref.~\cite{Bertacca:2007ux} and \cite{Gorini} where one impose the langrangian (i.e. the pressure) to be a constant, 
and also the model of  a single dark perfect fluid proposed in Ref.~\cite{Quercellini:2007ht}. For such cases it is clear that the 
low $\ell$ contribution~(\ref{I_theta_k-eta<1-2}) to the ISW effect will be the same that is predicted by the 
$\Lambda$CDM model. This is easily explained considering that for such long wevelenght perturbations the sound speed in fact plays 
no role at all.

\subsection{Derivation of $I_{\Theta_{l}}$ for modes $k \eta_{1/2} >1$}

As we have already mentioned in the previous section, in general a viable UDM must have a sound speed very close to zero 
for $\eta < \eta_{1/2}$ in order to behave as dark matter also at the perturbed level to form the structures we see today, 
and thus the gravitational potential will start to change in time for $\eta > \eta_{1/2}$. Therefore for the modes 
$k \eta_{1/2} >1$, in order to evaluate Eq.~(\ref{theta-approx}b) into Eq.~(\ref{I_theta_k-eta>1}) we can impose 
that $\eta_k > \eta_{1/2}$ which, from the definition of $\eta_k \simeq \eta_0 - (l+1/2)/k $, moves the lower limit of 
Eq.~(\ref{I_theta_k-eta>1}) to $(l+1/2)/(\eta_0-\eta_{1/2})$. 
Moreover we have that $\eta_{1/2}\sim \eta_0$. We can use this property
to estimate any observable at the value of $\eta_k$. Defining
\begin{equation}
\label{chikappadef}
\chi=\frac{\eta}{\eta_{1/2}}\, , \quad \quad {\rm and} \quad \quad \kappa=k \eta_{1/2}\, ,
\end{equation}
we have 
\begin{equation}
a_k=a(\eta_k)=a(\chi_k)=a_0+ \frac{da}{d\chi}\Bigg|_{\chi_0}\delta\chi_k
=1-\eta_{1/2}\mathcal{H}_0\frac{l+1/2}{\kappa}\, ,
\end{equation}
taking $a_0=1$, and 
\begin{equation}
\label{Phi_eta-k_approx}
\frac{d\Phi_{k}}{d\chi}(\chi_k)= \eta_{1/2} \Phi'(\eta_k)=
\frac{d\Phi_{k}}{d\chi}\Bigg|_{\chi_0} - \frac{d^2\Phi_{k}}{d\chi^2}
\Bigg|_{\chi_0}\left(\frac{l+1/2}{\kappa}\right)\; ,
\end{equation}
where $\delta\chi_k=\chi_k - \chi_0=(\eta_k - \eta_0)/\eta_{1/2}=-(l+1/2)/\kappa$. 
Notice that the expansion~(\ref{Phi_eta-k_approx}) is fully justified, since as already mentioned above, the mimimum value of $\kappa$ 
in Eq.~(\ref{I_theta_k-eta>1}) moves to $(l+1/2)/(\eta_0/\eta_{1/2}- 1)$, making $\delta\chi_k$ much less than 1. 
Therefore we can write 
\begin{eqnarray}
\label{Theta-approx}
\frac{\left|\Theta_{l\;\mathrm{ISW}}(\eta_0,k)\right|^2}{(2l+1)^2}&=&
4 \left|\frac{\Phi_k'(\eta_k) I_l}{k}\right|^2=\frac{4I_l^2}{\kappa^2}
\left|\frac{d\Phi_k}{d\chi}(\chi_k)\right|^2= \nonumber \\ 
&=& \frac{4I_l^2}{\kappa^2}\left[\left|\frac{d\Phi_{k}}{d\chi}(\chi_0)
\right|^2- 2 \frac{d\Phi_{k}}{d\chi}(\chi_0) \frac{d^2\Phi_{k}}{d\chi^2}
(\chi_0) \left(\frac{l+1/2}{\kappa}\right) +\right.\nonumber \\ 
&+& \left. \left|\frac{d^2\Phi_{k}}{d\chi^2}
(\chi_0)\right|^2\left(\frac{l+1/2}{\kappa}\right)^2\right]\;.
\end{eqnarray}

In this case, during $\eta_{1/2}<\eta<\eta_0$, there will be perturbation modes whose wavelength stays bigger than the Jeans 
length or smaller than it, i.e. we have to consider both the possibilities $c_s^2k^2 \ll \left|\theta''/\theta\right|$ and  
$c_s^2k^2 \gg \left|\theta''/\theta\right|$. In general the sound speed can vary with time, and in particular 
it might become significantly different from zero at late times. However, just as a first approximation, 
we exclude the intermediate situation because usually $\eta_{1/2}$ is very close to $\eta_0$ 
(this situation will be briefly analyzed later).\\

\subsubsection{Perturbation modes on scales bigger than the Jeans length. \\}
When $c_s^2k^2 \ll \left|\theta''/\theta\right|$ the value of $\Phi'(\eta_k)$ 
can be written from Eq.(\ref{Phi_k-eta<1}) as  
\begin{equation}
\label{phi'approx}
\Phi'(\eta_k)=\Phi_k(0) \tilde{\Phi}_k'(\eta_k)=\Phi_k(0) a(\eta_k) 
\left[\frac{d^2}{dt^2}\left(\frac{1}{a}
\int_{t_i}^{t}a(\tilde{t}) d\tilde{t}\right)\right]_{t=t(\eta_k)}\, .
\end{equation} 
Now, using this expression in Eq.~(\ref{Theta-approx}), with the primordial power spectrum $k^3 |\Phi_k(0)|^2=B k^{n-1}$, the value of 
(\ref{I_theta_k-eta>1}) may be written as 
\begin{eqnarray}
\label{I_theta_k-eta>1-cs_k<<theta''/theta}
\frac{I_{\Theta_{l}}(k \eta_{1/2} > 1)}{2l+1} &=& 4I_l^2 B \eta_{1/2}^{n-1}
\left[ \int_{\frac{l+1/2}{\chi_0 - 1}}^\infty \frac{d\kappa}{\kappa^3} 
\kappa^{n-1}  \left|\frac{d\tilde{\Phi}_k}{d\chi}(\chi_k)\right|^2\right]=
\nonumber \\ 
&=& 4I_l^2 B \eta_{1/2}^{n-1} \left[ 
\frac{1}{3-n} \left(\frac{\chi_0 - 1}{l+1/2}\right)^{3-n}
\left|\frac{d\tilde{\Phi}_{k}}{d\chi}(\chi_0) \right|^2  +\right.\nonumber \\
&-&\left.\frac{2(l+1/2)}{4-n} \left(\frac{\chi_0 - 1}{l+1/2}\right)^{4-n}
\frac{d\tilde{\Phi}_{k}}{d\chi}(\chi_0) \frac{d^2\tilde{\Phi}_{k}}{d\chi^2}
(\chi_0)+\right. \nonumber \\ 
&+& \left.\frac{(l+1/2)^2}{5-n}\left(\frac{\chi_0 - 1}{l+1/2}\right)^{5-n}
\left|\frac{d^2\tilde{\Phi}_{k}}{d\chi^2}
(\chi_0)\right|^2 \right]
\end{eqnarray}
with $(d\tilde{\Phi}_{k}/d\chi)_{\chi_0}=\eta_{1/2} \tilde{\Phi}_k'(\eta_0)$
and with $(d^2\tilde{\Phi}_{k}/d\chi^2)_{\chi_0}=
\eta_{1/2}^2\tilde{\Phi}_k''(\eta_0)$.\\
A second relevant comment follows from the fact that $I_l^2\sim 1/l$ for $l \gg 1$. We thus see that 
for $n=1$ and for $l \gg 1$ the contribution to the angular power spectrum from the modes under consideration is 
$l (l+1) C_l^{ISW}/(4\pi) = l(l+1)I_{\Theta_{l}}(k \eta_{1/2} > 1)/(2\pi^2(2l+1)) \sim 1/l$. 
In other words we find a similar slope as in \cite{Hu:1995em,Starobinsky} found in 
the $\mathrm{\Lambda CDM}$ model. Recalling the results of the previous section, this means that in the unified dark matter models 
the contribution to the ISW effect from those perturbations that are outside the Jeans length is very similar to the one produced in a 
$\Lambda$CDM model. The main difference on these scales will be present if the background evolution is different from the one in the 
$\Lambda$CDM model, but for the models where the background evolution is the same, as those proposed in Refs.~
\cite{Scherrer:2004au,Bertacca:2007ux,Gorini,Quercellini:2007ht} no difference at all can be observable.

\subsubsection{Perturbation modes on scales smaller than the Jeans length. \\}
When $c_s^2k^2 \gg \left|\theta''/\theta\right|$ we must use the solution  
(\ref{u-cs_k>>theta''/theta}) and through the relation~(\ref{u-v}a)  the gravitational potential is given by 
\begin{equation}
\label{usol}
\Phi_k(\eta)=\frac{1}{2}\left[(p+\rho)/c_s\right]^{1/2}(\eta) \mathit{C}_k(\eta_{1/2})
\cos\left(k\int_{\eta_{1/2}}^{\eta}c_s(\tilde{\eta})d\tilde{\eta} \right)\, .
\end{equation}
In Eq.~(\ref{usol}) $\mathit{C}_k(\eta_{1/2}) = \Phi_k(0)\mathit{C}_{1/2}$ is a constant of integration where 
\begin{equation}
\mathit{C}_{1/2}=2\frac{\left[1-\frac{\mathcal{H}(\eta_{1/2})}{a^2(\eta_{1/2})}
\left(I_R+\int_{\eta_{R}}^{\eta_{1/2}}a^2(\tilde{\eta})d\tilde{\eta}\right)
\right]}{\left[(p+\rho)/c_s\right]^{1/2}(\eta_{1/2})}\; ,
\end{equation}  
and it is obtained under the approximation that for $\eta < \eta_{1/2}$ one can use the longwavelenght solution~(\ref{Phi_k-eta<1}), 
since for these epochs
the sound speed must be very close to zero. Notice that Eq.~(\ref{usol}) shows clearly that the gravitational potential is oscillating 
and decaying in time. 

Defining for simplicity $\overline{C}^2 =\mathit{C}_{1/2}^2
[(p+\rho)/c_s](\eta_0)/4$, we take the time derivative of the gravitational potential appearing in Eq.~(\ref{theta-approx}b) by employing 
the expansion of Eq.(\ref{Theta-approx}). We thus find that Eq.~(\ref{I_theta_k-eta>1}) yields 
\begin{eqnarray}
\label{I_theta_k-eta>1-cs_k>>theta''/theta}
\fl
\frac{I_{\Theta_{l}}(k \eta_{1/2} > 1)}{2l+1}&=& 4 \overline{C}^2 
B I_l^2 \eta_{1/2}^{n-1} \left\{ \mathcal{C}_{\{k5,l2,c^2\}} (l+1/2)^2 
\left[\int_{\frac{l+1/2}{\chi_0 - 1}} ^\infty \frac{d\kappa}{\kappa^5} 
\kappa^{n-1} \cos^2(\mathit{D}_0 \kappa)\right]+ \right. \nonumber \\
&+& \mathcal{C}_{\{k4,l1,c^2\}} (l+1/2)\left[\int_{\frac{l+1/2}
{\chi_0 - 1}} ^\infty \frac{d\kappa}{\kappa^4} 
\kappa^{n-1} \cos^2(\mathit{D}_0 \kappa)\right] + \nonumber \\ 
&+&  \mathcal{C}_{\{k4,l2,sc\}} (l+1/2)^2 
\left[\int_{\frac{l+1/2}{\chi_0 - 1}} ^\infty \frac{d\kappa}{\kappa^4} 
\kappa^{n-1} \cos(\mathit{D}_0 \kappa)\sin(\mathit{D}_0 \kappa)\right]+
 \nonumber \\
&+& \left[ \mathcal{C}_{\{k3,l0,c^2\}} + \mathcal{C}_{\{k3,l2,c^2\}}(l+1/2)^2 
\right] \left[\int_{\frac{l+1/2}{\chi_0 - 1}} ^\infty \frac{d\kappa}{\kappa^3} 
\kappa^{n-1} \cos^2(\mathit{D}_0 \kappa)\right] + \nonumber \\ 
&+& \mathcal{C}_{\{k3,l2,s^2\}}(l+1/2)^2 \left[\int_{\frac{l+1/2}{\chi_0 - 1}}
^\infty \frac{d\kappa}{\kappa^3} \kappa^{n-1} \sin^2(\mathit{D}_0 \kappa)
\right] \nonumber \\
&+& \mathcal{C}_{\{k3,l1,sc\}}(l+1/2)\left[\int_{\frac{l+1/2}{\chi_0 - 1}} 
^\infty \frac{d\kappa}{\kappa^3} \kappa^{n-1} \cos(\mathit{D}_0 \kappa)
\sin(\mathit{D}_0 \kappa)\right]+\nonumber \\
&+& \mathcal{C}_{\{k2,l1,c^2\}}(l+1/2)\left[\int_{\frac{l+1/2}{\chi_0 - 1}} 
^\infty \frac{d\kappa}{\kappa^2} \kappa^{n-1} \cos^2(\mathit{D}_0 \kappa)
\right] + \nonumber \\ 
&+& \mathcal{C}_{\{k2,l1,s^2\}}(l+1/2)\left[\int_{\frac{l+1/2}{\chi_0 - 1}} 
^\infty \frac{d\kappa}{\kappa^2} \kappa^{n-1} \sin^2(\mathit{D}_0 \kappa)
\right] + \nonumber \\ 
&+& \left[ \mathcal{C}_{\{k2,l0,sc\}} + \mathcal{C}_{\{k2,l2,sc\}}(l+1/2)^2 
\right] \left[\int_{\frac{l+1/2}{\chi_0 - 1}} 
^\infty \frac{d\kappa}{\kappa^2} \kappa^{n-1} \cos(\mathit{D}_0 \kappa)
\sin(\mathit{D}_0 \kappa)\right]+\nonumber \\
&+& \mathcal{C}_{\{k1,l0,s^2\}}\left[\int_{\frac{l+1/2}{\chi_0 - 1}} 
^\infty \frac{d\kappa}{\kappa} \kappa^{n-1} \sin^2(\mathit{D}_0 \kappa)
\right] + \nonumber \\ 
&+& \mathcal{C}_{\{k1,l2,c^2\}}(l+1/2)^2\left[\int_{\frac{l+1/2}{\chi_0 - 1}} 
^\infty \frac{d\kappa}{\kappa} \kappa^{n-1} \cos^2(\mathit{D}_0 \kappa)
\right] + \nonumber \\ 
&+& \left.\mathcal{C}_{\{k1,l1,sc\}}(l+1/2)\left[\int_{\frac{l+1/2}
{\chi_0 - 1}} ^\infty \frac{d\kappa}{\kappa} \kappa^{n-1} \cos(\mathit{D}_0 \kappa)\sin(\mathit{D}_0 \kappa)\right]\right\}
\end{eqnarray}
with $\mathit{D}_0=\int_1^{\chi_0} c_s(\tilde{\chi})d\tilde{\chi}$
and where 
\begin{eqnarray}
\label{C{k,l}}
\fl 
\mathcal{C}_{\{k5,l2,c^2\}}=\left\{\frac{(p+\rho)_{,\chi\chi}}{(p+\rho)}
-\left(\frac{(p+\rho)_{,\chi}}{(p+\rho)}-\frac{c_{s,\chi}}{c_s}\right)
\left[2\frac{c_{s,\chi}}{c_s}+\frac{1}{2}\left(\frac{(p+\rho)_{,\chi}}
{(p+\rho)}-\frac{c_{s,\chi}}{c_s}\right)\right]\right\}^2
\Bigg|_{\chi_0}\;,\nonumber \\ 
\fl\mathcal{C}_{\{k4,l1,c^2\}}=-2\Bigg\{\left(\frac{(p+\rho)_{,\chi}}
{(p+\rho)}-\frac{c_{s,\chi}}{c_s}\right)^2
\left[\frac{(p+\rho)_{,\chi\chi}}{(p+\rho)}\right. \nonumber \\ 
 \left. -\left(\frac{(p+\rho)_{,\chi}}{(p+\rho)}
-\frac{c_{s,\chi}}{c_s}\right)
\left(2\frac{c_{s,\chi}}{c_s}+\frac{1}{2}\left(\frac{(p+\rho)_{,\chi}}
{(p+\rho)}-\frac{c_{s,\chi}}{c_s}\right)\right)\right]
\Bigg\}\Bigg|_{\chi_0}\;,\nonumber \\ 
\fl \mathcal{C}_{\{k4,l2,sc\}}=4\left\{
c_s \frac{(p+\rho)_{,\chi}}{(p+\rho)} 
\left[\frac{(p+\rho)_{,\chi\chi}}{(p+\rho)}
-\left(\frac{(p+\rho)_{,\chi}}{(p+\rho)}-\frac{c_{s,\chi}}{c_s}\right)
\right. \right.\nonumber \\
\left.\left.\left(2\frac{c_{s,\chi}}{c_s}+\frac{1}{2}
\left(\frac{(p+\rho)_{,\chi}}
{(p+\rho)}-\frac{c_{s,\chi}}{c_s}\right)\right) \right]\right\}
\Bigg|_{\chi_0}\;,\nonumber \\ 
\fl \mathcal{C}_{\{k3,l0,c^2\}}=\left[\frac{(p+\rho)_{,\chi}}{(p+\rho)}
-\frac{c_{s,\chi}}{c_s}\right]^2\Bigg|_{\chi_0}\;, \quad \quad \quad
\mathcal{C}_{\{k3,l2,s^2\}}=4\left[c_s\frac{(p+\rho)_{,\chi}}
{(p+\rho)}\right]^2\Bigg|_{\chi_0}\;,
\nonumber \\ 
\fl \mathcal{C}_{\{k3,l2,c^2\}}=4
\left\{c_s^2\left[\frac{(p+\rho)_{,\chi\chi}}{(p+\rho)}
-\left(\frac{(p+\rho)_{,\chi}}{(p+\rho)}-\frac{c_{s,\chi}}{c_s}\right)
\left(2\frac{c_{s,\chi}}{c_s}+\frac{1}{2}\left(\frac{(p+\rho)_{,\chi}}
{(p+\rho)}-\frac{c_{s,\chi}}{c_s}\right)\right) \right]
\right\}\Bigg|_{\chi_0}\;,\nonumber \\ 
\fl\mathcal{C}_{\{k3,l1,sc\}}=4\left\{c_s
\left[\frac{(p+\rho)_{,\chi\chi}}{(p+\rho)}
-\left(\frac{(p+\rho)_{,\chi}}{(p+\rho)}-\frac{c_{s,\chi}}{c_s}\right)
\left(2\frac{c_{s,\chi}}{c_s}+\frac{1}{2}\left(\frac{(p+\rho)_{,\chi}}
{(p+\rho)}-\frac{c_{s,\chi}}{c_s}\right)\right) \right]\right. \nonumber \\ 
\left.-c_s\frac{(p+\rho)_{,\chi}}{(p+\rho)}\left[\frac{(p+\rho)_{,\chi}}
{(p+\rho)}-\frac{c_{s,\chi}}{c_s}\right]\right\}\Bigg|_{\chi_0}\;,\nonumber \\
\fl \mathcal{C}_{\{k2,l1,c^2\}}=-4\left\{c_s^2\left[\frac{(p+\rho)_{,\chi}}
{(p+\rho)}-\frac{c_{s,\chi}}{c_s}\right]\right\}\Bigg|_{\chi_0}
\;, \quad \quad \quad 
\mathcal{C}_{\{k2,l1,s^2\}}=8\left[c_s^2 \frac{(p+\rho)_{,\chi}}
{(p+\rho)}\right]\Bigg|_{\chi_0}\;,\nonumber\\
\fl \mathcal{C}_{\{k2,l0,sc\}}=-4\left\{c_s\left[\frac{(p+\rho)_{,\chi}}
{(p+\rho)}-\frac{c_{s,\chi}}{c_s}\right]\right\}\Bigg|_{\chi_0}
\;, \quad \quad \quad 
\mathcal{C}_{\{k2,l2,sc\}}=8\left[c_s^3 \frac{(p+\rho)_{,\chi}}
{(p+\rho)}\right]\Bigg|_{\chi_0}\;,\nonumber\\
\fl \mathcal{C}_{\{k1,l0,s^2\}}=4c_s^2\big|_{\chi_0}\;, \quad \quad  \quad 
\mathcal{C}_{\{k1,l2,c^2\}}=4c_s^4\big|_{\chi_0}\;, \quad \quad \quad  
\mathcal{C}_{\{k1,l1,sc\}}=8c_s^3\big|_{\chi_0}\;.
\end{eqnarray}

In this case we have defined $(\cdot)_{,\chi}\equiv d(\cdot)/d\chi$ and we recall that the dimensionless 
variables $\chi$ and $\kappa$ are defined in Eq.~(\ref{chikappadef}). We have indicated the coefficients 
${\mathcal C}_{\{k[j],l[i],[sc]\}}$ 
in such a way to signal that they multiply an integral in $\kappa$ of $\kappa^{n-1}/\kappa^j$ times $\sin(D_0\kappa)\cos(D_0\kappa)$ and the overall 
multipole coefficent is $(l+1/2)^i$. We can infer from (\ref{I_theta_k-eta>1-cs_k>>theta''/theta})
that for $n<1$ all integrals are convergent. Notice that a natural cut-off in the various integrals is introduced for those modes that enter the 
horizon during the radiation dominated epoch, due to the Meszaros effect that the matter fluctuations will suffer until the full matter domination epoch. 
Such a cut-off will show up in the gravitational potential and in the various integrals of  Eq.~(\ref{I_theta_k-eta>1-cs_k>>theta''/theta}) as a 
$(k_{eq}/k)^4$ factor, where $k_{eq}$ is the wavenumber of the Hubble radius at the equality epoch. 

A simple inspection of Eq.~(\ref{I_theta_k-eta>1-cs_k>>theta''/theta}) 
shows one of our main results. The terms of Eq.~(\ref{I_theta_k-eta>1-cs_k>>theta''/theta}) where the coefficients 
$\mathcal{C}$ turn out to be proportional to the sound speed $c_s$ cause the growth of $l(l+1)I_{\Theta_{l}}(k \eta_{1/2} > 1)/(2l+1)$, 
(and hence of the power spectrum $l(l+1) C_l$ through Eq.~(\ref{Cl2})), as $l$ increases. 
This means that, if the sound speed of the unified dark matter fluid starts to 
differ significantly from zero at late times, the consequence is to produce a very strong ISW effect, and clearly this does not happen in 
a $\Lambda$CDM universe since $c_s^2=0$ always. This effect is easily explained by considering that 
the energy density of the universe in the unified models is dominated at late time by a 
just single fluid. Therefore an eventual appearance of a Jeans lenght (i.e. a departure of the sound speed from zero) makes 
the oscillating behaviour of the dark fluid pertubations under the Jeans lenght immediately visible through a strong time dependence of the 
gravitational potential. In fact one can verify that the scalar field fluctuations~(\ref{deltaphi}) are oscillating and decaying in time as 
$\delta \varphi \sim (k/a)\left[c_s \big/ (\partial p/\partial X)\right]^{1/2}\sin(k\int_{\eta_{1/2}}^\eta c_s d\eta)$. Similar results have been discussed in the case of the GCG model in 
Refs.~\cite{Carturan:2002si,Amendola:2003bz}.

We point out that the potentially most dangerous term in Eq.~(\ref{I_theta_k-eta>1-cs_k>>theta''/theta}) 
is the one identified by the coefficient ${\mathcal C}_{{\{k1,l2,c^2}\}}$ 
\begin{equation}
4c_s^4\big|_{\chi_0}\,(l+1/2)^2\left[\int_{\frac{l+1/2}{\chi_0 - 1}} 
^\infty \frac{d\kappa}{\kappa} \kappa^{n-1} \cos^2(\mathit{D}_0 \kappa)
\right]\, .
\end{equation}
Such a term makes the power spectrum $l(l+1)C_l$ to scale as $l^3$ 
until $l\approx 25$. This angular scale is obtained by considering the peak of the Bessel functions in correspondence of the cut-off scale 
$k_{eq}$, 
$l \approx k_{eq}(\eta_0-\eta_{1/2})$. In fact, for smaller scales, the integral identified by the coefficient 
${\mathcal C}_{{\{k1,l2,c^2}\}}$ will 
decrease as $1/ \ell$.

\subsubsection{Intermediate case.\\} 
Now we shall briefly discuss the intermediate case that corresponds to perturbation modes that initially are outside the Jeans 
length and then, 
due to a time variation of the sound speed, they fall inside them. This corresponds to consider the range 
$[(l+1/2)/(\chi_0 - 1)]^2 < \kappa^2_J = \left|\theta_{,\chi\chi}/\theta\right|/c_s^2 < 
\kappa_{eq}^2$. In this case $\kappa > (l+1/2)/(\chi_0 - 1)$
and so we can use the same procedure described before. 
Indeed when $k \sim k_J$ i.e. $c_s^2 k_J^2 \sim \left|\theta''/\theta\right|$ 
it can be written as follows 
\begin{equation}
\left\{\left[ c_s^2 \left|\theta_{,\chi\chi}/\theta\right|^{-1}\right]
\bigg|_{\chi_0}-\left[ c_s^2 \left|\theta_{,\chi\chi}/\theta\right|^{-1}\right]
_{,\chi}\bigg|_{\chi_0}\left(\frac{l+1/2}{\kappa_J}\right)\right\}\kappa_J^2 =1
\end{equation}
i.e.
\begin{equation}
\kappa_J=\kappa_J(l)=B_l+\sqrt{B_l^2+A}
\end{equation}
where $B_l\equiv\{[\ln(c_s^2)]_{,\chi}-
[\ln\left|\theta_{,\chi\chi}/\theta\right|]_{,\chi}\}\big|_{\chi_0}(2l+1)/4$ 
and $A\equiv[\left|\theta_{,\chi\chi}/\theta\right|/(2c_s^2)]\big|_{\chi_0}$.
We immediately see that $I_{\Theta_{l}}(k \eta_{1/2} > 1)$ can be divided into 
 two parts. The first part is identical to 
(\ref{I_theta_k-eta>1-cs_k<<theta''/theta}) except for the upper limit of 
the integral. Indeed now the upper limit is $\kappa_J(l)$. 
In order to derive the second part we note that now the lower limit of the
integral in $\kappa$ is $\kappa_J(l)$ and that $u_k(\eta_k)=
[\mathit{C}_k(\eta_J)/c_s^{1/2}(\eta_k)]
\cos\left(k\int_{\eta_J}^{\eta_k}c_s(\tilde{\eta})d\tilde
{\eta} \right)$ where $\eta_J$ is the conformal time when 
$c_s^2\big|_{\eta_J} k^2 \sim \left|\theta''/\theta\right|_{\eta_J}$ 
($\eta_J$ is function of $k$) and 
\begin{equation}
\mathit{C}_k(\eta_J) = 2 \Phi_k(0)
\frac{\left[1-\frac{\mathcal{H}(\eta_J)}{a^2(\eta_J)}
\left(I_R+\int_{\eta_{R}}^{\eta_J}a^2(\tilde{\eta})d\tilde{\eta}\right)
\right]}{[(p+\rho)/c_s]^{1/2}(\eta_J)}\;.
\end{equation}  

\section{Discussion of some examples and conclusions}

In most of the UDM models there are several properties in common. 
It is easy to see that in Eq.(\ref{Ir}) $I_R$  is negligible because of the low value
of $a_{eq}$.\\
Moreover in the various models usually we have that strong differences with respect to the ISW effect in the $\Lambda$CDM case
can be produced from those scales that are inside the Jeans length as the photons pass through them. For these scales 
(which depend on the particular model) the perturbations of the unified dark matter fluid play the main role. On larger scales instead we 
find that they play no role and ISW signatures different from the $\Lambda$CDM case can come only from the different background expansion history.   
We have found that when $k^2 \gg k^2_J=c_s^{-2}
\left|\theta''/\theta\right|$ (see~(\ref{diff-eq_u})) one must take care of  the term in  
Eq.~(\ref{I_theta_k-eta>1-cs_k>>theta''/theta}) proportional to 
$\mathcal{C}_{\{k1,l2,c^2\}}$. Indeed this term grows faster than
the other integrals contained in (\ref{I_theta_k-eta>1-cs_k>>theta''/theta})
when $l$ increases up to $l\approx 25$. It is responsible for a strong ISW effect and hence it will cause in the CMB power spectrum 
$l(l+1)C_l/(2\pi)$ a decrease in the peak to plateau ratio (once the CMB power spectrum is normalized).  
In order to avoid this effect, a sufficient (but not necessary) condition
is that all the models have to satisfy $c_s^2k^2 < \left|\theta''
/\theta\right|$ for the scales of interest. The maximum constraint is found in correspondence of the scale at which the contribution 
Eq.~(\ref{I_theta_k-eta>1-cs_k>>theta''/theta}) proportional to $\mathcal{C}_{\{k1,l2,c^2\}}$ takes it maximum value, that is 
$k \approx k_{eq}$. For example in the 
Generalized Chaplygin Gas model (GCG), i.e when 
$p=-\Lambda^{1/(1+\alpha)}/\rho^\alpha$ and $c_s^2=-\alpha w$, we deduce that 
$|\alpha| <  10^{-4}$ (see Refs.~\cite{Bento:2002ps}
\cite{Carturan:2002si} \cite{Amendola:2003bz}
and \cite{GCG}). This is also in accordance with 
\cite{Sandvik:2002jz} which performs an analysis on the mass power spectrum and gravitational lensing constraints thus finding a more 
stringnet 
constraint. \\
As far as the generalized  Scherrer solution models \cite{Bertacca:2007ux} are concerned, in these models the pressure of the unified 
dark matter 
fluid is given by $p=g_n (X-X_0)^n-\Lambda$, where $g_n$ is a suitable constant and $n>1$. 
The case $n=2$ corresponds to unified model proposed by Scherrer~\cite{Scherrer:2004au}. In this case we find that imposing the 
constraint $c_s^2k^2 < \left|\theta''/\theta\right|$ for the scales of interest we get that 
$\epsilon=(X-X_0)/X_0 < (n-1)\;10^{-4}$.\\
If we want to study in greater detail what happens in the GCG model
when $c_s^2k^2 \gg \left|\theta''/\theta\right|$ we discover the following 
things:
\begin{itemize}
\item
for $10^{-4}<\alpha \leq 5 \times 10^{-3}$, where  we
are in the ``Intermediate case''. Now $c_s^2=-\alpha w$ is very small
and the background of the cosmic expansion history of the 
universe is very similar to the $\mathrm{\Lambda CDM}$ model.
In this situation the pathologies, described before, 
are completely negligible.
\item
When treating $ 6 \times 10^{-3} < \alpha \leq 1$ a very strong ISW effect will be produced and we have estimated the same orders of 
magnitude for the decrease of the peak to plateau ratio in the anisotropy spectrum  
$l(l+1)C_l/(2\pi)$ (once it is normalized) that can be inferred from the authors of \cite{Carturan:2002si} 
obtained in the numerical simulations (having assumed that the production of the peaks during the acosutic oscillations at
 recombination is similar 
to what happens in a $\Lambda$CDM model, since at recombination the effects of the sound speed will be negligible).\\
\end{itemize}
An important observation arises when considering those UDM models
that reproduce the same cosmic expansion history of the 
universe as the $\mathrm{\Lambda CDM}$ model. Among these models one can impose the condition  
$w=-c_s^2$ which, for example, is predicted by UDM models with a kinetic term of the Born-Infeld type 
\cite{Bertacca:2007ux} \cite{Gorini} \cite{Padmanabhan:2002sh}. In this case, computing the integral 
in Eq.~(\ref{I_theta_k-eta>1-cs_k>>theta''/theta}) proportional to $\mathcal{C}_{\{k1,l2,c^2\}}$ which give the main contribution 
to the ISW effect 
we have estimated that the corresponding decrease of peak to plateau ratio is about
one third with respect to what we have in the GCG when the value of $\alpha$ 
is equal to 1. The special case $\alpha=1$ is called  ``Chaplygin Gas'' 
(see for example \cite{Bilic02}) and it is characterized by a background equation of state $w$ which evolves in a different way 
to the standard $\Lambda$CDM case. \\ 
From these considerations we deduce that this specific effect stems
only in part from  the background of the cosmic expansion history of the 
universe and that the most relevant contribution to the ISW effect is due to the value of the speed of sound $c_s^2$.

\section*{Acknowledgments}
We thank Sabino Matarrese and Massimo Pietroni for useful discussions.



\section*{References}

\begin{thebibliography}{99}

\bibitem{Weinberg:1988cp}
S.~Weinberg,
Rev.\ Mod.\ Phys.\  {\bf 61}, 1 (1989).

\bibitem{perlmutter} 
S.~Perlmutter {\it et al.},
Astrophys.\ J.\  {\bf 517}, 565 (1999).
\bibitem{riess} 
A.~G.~Riess {\it et al.},
Astron.\ J.\  {\bf 116}, 1009 (1998); {\it idem},  
Astron.\ J.\  {\bf 117}, 707 (1999).


\bibitem{Wetterich88}
C.~Wetterich, Nucl. \ Phys \ B. {\bf 302}, 668 (1988).

\bibitem{peebles}
B.~Ratra and J.~Peebles, Phys. \ Rev \ D {\bf 37}, 321 (1988).
 
\bibitem{earlyqu}
Y.~Fujii,
Phys.\ Rev.\ D {\bf 26}, 2580 (1982);
L.~H.~Ford,
Phys.\ Rev.\ D {\bf 35}, 2339 (1987);
Y.~Fujii and T.~Nishioka,
Phys.\ Rev.\ D {\bf 42}, 361 (1990);
T.~Chiba, N.~Sugiyama and T.~Nakamura,
Mon.\ Not.\ Roy.\ Astron.\ Soc.\  {\bf 289}, L5 (1997).

\bibitem{Carroll:1998zi}
S.~M.~Carroll,
Phys.\ Rev.\ Lett.\  {\bf 81}, 3067 (1998).

\bibitem{Ferreira97}
P.~G.~Ferreira and M.~Joyce,
Phys.\ Rev.\ Lett.\  {\bf 79}, 4740 (1997); {\it idem}
Phys.\ Rev.\ D {\bf 58}, 023503 (1998).

\bibitem{CLW}
  E.~J.~Copeland, A.~R.~Liddle and D.~Wands,
  Phys.\ Rev.\ D {\bf 57}, 4686 (1998).

\bibitem{caldwell98} 
R.~R.~Caldwell, R.~Dave and P.~J.~Steinhardt,
Phys.\ Rev.\ Lett.\  {\bf 80}, 1582 (1998).

\bibitem{Paul99} 
I.~Zlatev, L.~M.~Wang and P.~J.~Steinhardt,
Phys.\ Rev.\ Lett.\  {\bf 82}, 896 (1999);
P.~J.~Steinhardt, L.~M.~Wang and I.~Zlatev, 
Phys.\ Rev.\ D {\bf 59}, 123504 (1999).


\bibitem{Chiba:1999ka}  T.~Chiba, T.~Okabe and M.~Yamaguchi, 
Phys.\ Rev.\ D {\bf 62}, 023511 (2000).  


\bibitem{AMS1} 
C. Armend\'ariz-Pic\'on, V. Mukhanov, and P. J. Steinhardt, 
Phys. \ Rev. \ Lett. {\bf 85}, 4438 (2000).

\bibitem{AMS2} 
C. Armend\'ariz-Pic\'on, V. Mukhanov, and P. J. Steinhardt, 
Phys. \ Rev. \ D {\bf 63}, 103510 (2001).



\bibitem{kinf} 
C. Armend\'ariz-Pic\'on, T. Damour, and V. Mukhanov, 
Phys. \ Lett. \ B {\bf 458}, (1999) 209;

\bibitem{Garriga:1999vw}
  J.~Garriga and V.~F.~Mukhanov,
  Phys.\ Lett.\  B {\bf 458} (1999) 219
  [arXiv:hep-th/9904176].


\bibitem{Copeland:2006wr}
  E.~J.~Copeland, M.~Sami and S.~Tsujikawa,
  Int.\ J.\ Mod.\ Phys.\  D {\bf 15}, 1753 (2006). 






\bibitem{Kamenshchik:2001cp}
A.~Y.~Kamenshchik, U.~Moschella and V.~Pasquier,
Phys.\ Lett.\ B {\bf 511}, 265 (2001).

\bibitem{Bilic02}
N.~Bilic, G.~B.~Tupper and R.~D.~Viollier,
Phys.\ Lett.\ B {\bf 535}, 17 (2002).

\bibitem{Bento:2002ps}
M.~C.~Bento, O.~Bertolami and A.~A.~Sen,
Phys.\ Rev.\ D {\bf 66}, 043507 (2002).

\bibitem{Scherrer:2004au}
  R.~J.~Scherrer,
  Phys.\ Rev.\ Lett.\  {\bf 93} (2004) 011301

\bibitem{Bertacca:2007ux}
  D.~Bertacca, S.~Matarrese and M.~Pietroni,
  arXiv:astro-ph/0703259, accepted for publication in Mod. Phys. Lett. A.

\bibitem{Quercellini:2007ht}
  C.~Quercellini, M.~Bruni and A.~Balbi,
  arXiv:0706.3667 [astro-ph].

\bibitem{DiezTejedor:2006qh}
  A.~Diez-Tejedor and A.~Feinstein,
  Phys.\ Rev.\  D {\bf 74} (2006) 023530

\bibitem{Gorini}
  V.~Gorini, A.~Y.~Kamenshchik, U.~Moschella and V.~Pasquier,
  Phys.\ Rev.\  D {\bf 69} (2004) 123512,

V.~Gorini, A.~Kamenshchik, U.~Moschella, V.~Pasquier and A.~Starobinsky,
  Phys.\ Rev.\  D {\bf 72} (2005) 103518

\bibitem{Padmanabhan:2002sh}
  T.~Padmanabhan and T.~R.~Choudhury,
  Phys.\ Rev.\  D {\bf 66} (2002) 081301

\bibitem{takayana}
  F.~Takahashi and T.~T.~Yanagida,
  Phys.\ Lett.\  B {\bf 635}, 57 (2006). 



\bibitem{bono}
 R.~Mainini and S.~A.~Bonometto,
  Phys.\ Rev.\ Lett.\  {\bf 93}, 121301 (2004)



\bibitem{WL}
  J.~Weller and A.~M.~Lewis,
  Mon.\ Not.\ Roy.\ Astron.\ Soc.\  {\bf 346}, 987 (2003).
 




\bibitem{DD}
  S.~DeDeo, R.~R.~Caldwell and P.~J.~Steinhardt,
  Phys.\ Rev.\  D {\bf 67}, 103509 (2003)
  [Erratum-ibid.\  D {\bf 69}, 129902 (2004)].
  


\bibitem{BO}
  R.~Bean and O.~Dore,
  Phys.\ Rev.\  D {\bf 69}, 083503 (2004).



\bibitem{Huss}
  W.~Hu,
  Astrophys.\ J.\  {\bf 506}, 485 (1998).

  




\bibitem{BDG}
  R.~Bean and O.~Dore,
  Phys.\ Rev.\  D {\bf 68}, 023515 (2003).



\bibitem{Carturan:2002si}
  D.~Carturan and F.~Finelli,
  Phys.\ Rev.\  D {\bf 68} (2003) 103501



\bibitem{Amendola:2003bz}
  L.~Amendola, F.~Finelli, C.~Burigana and D.~Carturan,
  JCAP {\bf 0307} (2003) 005
  [



\bibitem{Sandvik:2002jz}
H.~Sandvik, M.~Tegmark, M.~Zaldarriaga and I.~Waga,
  Phys.\ Rev.\  D {\bf 69} (2004) 123524


\bibitem{Giannakis-Hu} 
D. Giannakis and W. Hu, Phys. Rev. D {\bf 72}, 063502 (2005).

\bibitem{Sachs:1967er}
  R.~K.~Sachs and A.~M.~Wolfe,
  Astrophys.\ J.\  {\bf 147} (1967) 73.

\bibitem{Hu:1995em}
  W.~T.~Hu,
  ``Wandering in the background: A Cosmic microwave background explorer'' PhD Thesis, 
  arXiv:astro-ph/9508126.



\bibitem{Mukhanov:1990me}
  V.~F.~Mukhanov, H.~A.~Feldman and R.~H.~Brandenberger,
  Phys.\ Rept.\  {\bf 215} (1992) 203.


\bibitem{Mukhanov:2005sc}
  V.~Mukhanov, ``Physical foundations of cosmology'',
  {\it  Cambridge, UK: Univ. Pr. (2005) 421 p}

\bibitem{Starobinsky}
 L.A.~Kofman and A.A.~Starobinsky,
 (Sov.) Astron. Lett. {\bf 11} (1985) 271 


\bibitem{GCG}

M.~Makler, S.~Quinet de Oliveira and I.~Waga,
  Phys.\ Lett.\  B {\bf 555} (2003) 1,
\\
M.~d.~C.~Bento, O.~Bertolami and A.~A.~Sen,
  Phys.\ Rev.\  D {\bf 67} (2003) 063003,
\\
J.~S.~Alcaniz, D.~Jain and A.~Dev,
  Phys.\ Rev.\  D {\bf 67} (2003) 043514
  [arXiv:astro-ph/0210476].
\\
M.~d.~C.~Bento, O.~Bertolami and A.~A.~Sen,
  Phys.\ Lett.\  B {\bf 575} (2003) 172, 

\end{thebibliography}
\end{document}